\documentclass[11pt]{article}
\usepackage{cite}
\usepackage{amssymb,amsmath,amsthm,mathrsfs}
\newcommand{\ot}{{\,\otimes\,}}
\def\oper{{\mathchoice{\rm 1\mskip-4mu l}{\rm 1\mskip-4mu l}
{\rm 1\mskip-4.5mu l}{\rm 1\mskip-5mu l}}}
\usepackage[english]{babel}
\usepackage{graphicx}
\usepackage{dcolumn}
\usepackage{bm}
\usepackage{color}
\usepackage{subfigure}
\usepackage[ansinew]{inputenc}
\usepackage{amsfonts}
\usepackage{amsmath}
\def\<{\langle}
\def\>{\rangle}

\def\e{\mathrm{e}}

\def\e{e}

\newtheorem{Theorem}{Theorem}

\newtheorem{Remark}{Remark}

\begin{document}

\title{\bf A Brief History of the GKLS Equation}
\author{Dariusz Chru\'sci\'nski \\
Institute of Physics, Faculty of Physics, Astronomy and Informatics \\
Nicolaus Copernicus University, Grudzi\c{a}dzka 5/7, 87--100 Toru\'n, Poland\\ [1ex]
email: darch@fizyka.umk.pl \\ \\
Saverio Pascazio \\
Dipartimento di Fisica and MECENAS, Universit\`{a} di Bari, I-70126 Bari, Italy \\
Istituto Nazionale di Ottica (INO-CNR), I-50125 Firenze, Italy \\
INFN, Sezione di Bari, I-70126 Bari, Italy\\[1ex]
e-mail: saverio.pascazio@ba.infn.it}


\maketitle

\abstract{We reconstruct the chain of events, intuitions and ideas that led to the formulation of the Gorini, Kossakowski, Lindblad and Sudarshan equation.}


\section{Introduction}
\label{intro}
\begin{quotation}

\noindent
{\it If you take any two historical events and you ask whether there are similarities and differences, the answer is always going to be both ``yes" and ``no." At some sufficiently fine level of detail there will be
differences, and at some sufficiently abstract level there will be similarities. The question we want to ask in the two cases we are considering, [$\ldots$] is whether the level at which there are similarities is, in fact, a significant one.}  \cite{Chomsky}
\end{quotation}%
The articles written by Vittorio Gorini, Andrzej Kossakowski and George Sudarshan (GKS) \cite{GKS}, and G\"oran Lindblad \cite{Lindblad} belong to the list of the most influential papers in theoretical physics. They were
published almost at the same time: the former \cite{GKS} in May 1976 and the latter \cite{Lindblad} in June 1976. Interestingly, they were also submitted simultaneously: \cite{GKS} on March 19th, 1975, and \cite{Lindblad} on April 7th, 1975. Archive and on-line submission did not exist in the '70s, so both papers were very probably in gestation at the same time.

It is always difficult to reconstruct facts from (necessarily incomplete) data. This is the work of historians. Things are even more complicated when one has to reconstruct how \emph{ideas} are born, and when and where.
We decided to resort to the wisdom of Noam Chomsky \cite{Chomsky} and analyze events, as well as detailed technical findings. However, while writing this note, we realized that one cannot just compare two articles or (worse) two equations: after all, GKS and Lindblad derived and published a very similar equation at the same time. It is more interesting to compare hypotheses and derivation, and the way the necessary concepts are formulated. Ideas are born in complicated fashions. Environments, influences, conferences, discussions, scientific literature are but a number of factors that play a crucial role.

Michel Berry, in his beautiful webpage, quotes Andr\'e Gide:
\begin{quotation}
\noindent
{\it Everything has been said before, but since nobody listens we have to keep going back and beginning all over again.}\footnote{Toutes choses sont dites d\'ej\`a; mais comme personne n'\'ecoute, il faut toujours recommencer. (French can be more concise, once in a while.)}
\end{quotation}
True. But after all this article must be kept finite, and we shall abstain from going back to Aristotle. The only way out of this dilemma is to try and define Chomsky's ``level of detail at which there are similarities" and decide whether it is a significant one. This will entail a certain degree of arbitrariness. We ask our readers to make allowance for our superficiality.

Let us mention a few facts.
In 1972, Andrzej Kossakowski published a largely un-noticed article \cite{Kossa1972} in which he proposed an axiomatic definition of dynamical semigroup.
From March 26th to April 6th, 1973, Vittorio Gorini attended the conference ``Foundations of quantum mechanics and ordered linear spaces", in Marburg, where he listened to talks by E.\ St{\o}rmer and  K.\ Kraus, who both mentioned the notion of complete positivity.
From September to December 1974 Vittorio Gorini and Andrzej Kossakowski were both visiting George Sudarshan, at the University of Texas at Austin.\footnote{Gorini had already visited Sudarshan in Texas from Fall 1971 to Spring 1973, but they were working on the classification of \emph{positive} (not completely positive) maps.}
G\"oran Lindblad used to work alone. He defended his thesis in May 1974. In December 1974 he participated in the Symposium on Mathematical Physics, organized in Toru\'n by Roman Ingarden --- Kossakowski's PhD supervisor. He sent an interesting recollection letter of those times to the organizers of the 48th SMP meeting. This letter is reproduced in Sect.\ \ref{Lrec}.
In January 1975 Gorini went to visit Lindblad in Stockholm, on the way back from Texas. Kossakowski never met Lindblad.
The two articles that concern us here \cite{GKS,Lindblad} were submitted between March and April 1975.
In 1980 Kraus spent a sabbatical year at the University of Texas at Austin; besides George Sudarshan, there were John Archibald Wheeler, Arno B\"ohm and William Wootters.

\begin{figure}
\centering
\includegraphics[width=129mm]{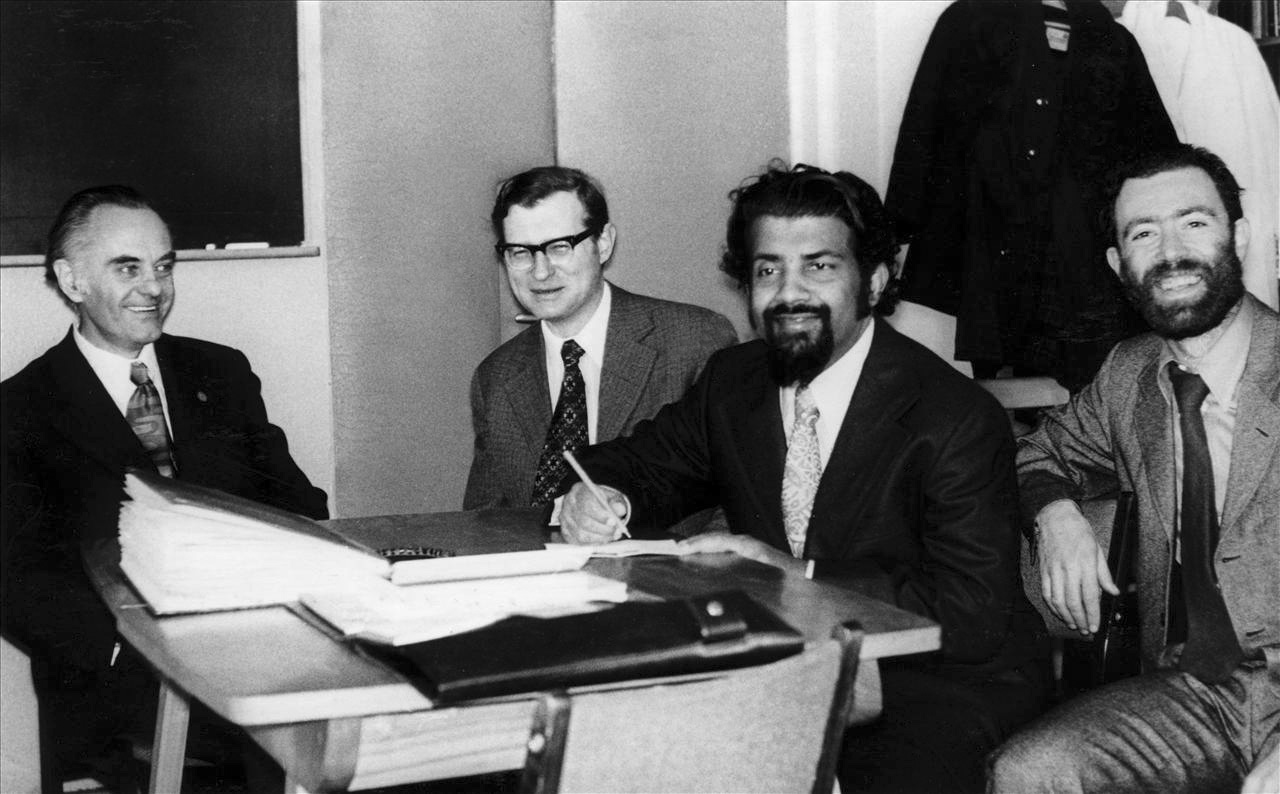}
\caption{\small Picture taken in Prof.\ Ingarden's office (December 1975). From left to right: Roman Ingarden, Andrzej Kossakowski, George Sudarshan and Vittorio Gorini.}
\label{fig:GKSI}
\end{figure}

We shall consider these facts in the following. Now let us clarify what this article is \emph{not}. It is not a review paper. It is not tutorial. It does not contain novel results. It does assume some familiarity with dissipative quantum systems and notation. It makes full use of our personal interactions with the four protagonists of this story.
Figure \ref{fig:GKSI} shows a 1975 picture taken in Torun, Poland, in Roman Ingarden's office.

\section{Formulation of the Problem}
\label{tut}

The evolution of a closed (isolated) quantum system is described by the Schr\"odinger equation (here and henceforth, Planck's constant $\hbar = 1$)
\begin{equation}
\label{schr}
i \dot \psi \;=\; H \psi \quad \longleftrightarrow \quad \psi_t \;=\; U_t \psi_0 \,,
\end{equation}
where $\psi$ is the wave function, $H$ the Hamiltonian, the dot denotes time derivative and $U_t= \e^{-i H t}$ is a one-parameter group. This translates into the so-called von Neumann equation for the density matrix $\rho$
\begin{equation}
\label{vn}
\dot \rho \;=\; -i [H,\rho] \quad \longleftrightarrow \quad \rho_t \;=\; U_t \rho_0 U^\dagger_t\,.
\end{equation}
If the quantum system is ``open", namely not isolated, and immersed in an environment with which it interacts, the above equation is not valid and must be replaced by the following evolution law
\begin{equation}
\label{superoperator}
\rho' \;=\; \Lambda \rho \;=\; \sum_\alpha K_\alpha  \rho K_\alpha^\dagger ,
\end{equation}
where $\sum_\alpha K_\alpha^\dagger  K_\alpha = \mathbb{I}$. Notice that if the summation is made up of a single addendum $K = U = \e^{-i H t}$, (\ref{superoperator}) reduces to (\ref{vn}). Notice also that (\ref{superoperator}) is not a differential equation, but rather takes a ``snapshot" of the quantum state at a particular time $t'$: $\rho' = \rho_{t'}$.

Equation (\ref{superoperator}) defines a quantum channel and $\Lambda$ is usually called a superoperator. It is assumed to be linear, it must preserve trace and hermiticity, and it must be completely positive. We shall delve into these properties in the following.

In the Markovian approximation, (\ref{superoperator}) yields the following differential
(``master") equation
\begin{equation}
\label{master}
\dot \rho \;=\;  \mathcal{L}\rho   \quad \longleftrightarrow \quad
\rho_t \;=\; e^{t \mathcal{L}} \rho_0\,,
\end{equation}
where $\mathcal{L}$ is the generator of a (one-parameter) quantum dynamical semigroup. The evolution (\ref{master}) inherits from (\ref{superoperator}) linearity, trace- and hermiticity-preservation, and complete positivity.

Our focus will be on the mathematical structure of (\ref{master}), but also on (\ref{superoperator}).
Physicists did not pay much attention to open quantum systems until rather recently, when, with the advent of the quantum era, topics such as quantum information and quantum technologies became crucial for the foundations and applications of quantum mechanics.

\section{The Structure of the GKLS Generator}
\label{equations}

The central problem addressed by Gorini, Kossakowski, Sudarshan and Lindblad (GKLS) \cite{GKS,Lindblad} was the characterization of the generator $\mathcal{L}$ of a quantum dynamical semigroup.
Consider quantum states, represented by density operators $\rho \in  \mathcal{T}_+(\mathcal{H})$ (positive trace-class operators) with $||\rho||_1 = {\rm Tr}\rho=1$.  The axioms for a dynamical semigroup had been elaborated by Kossakowski in 1972 \cite{Kossa1972} (see also  \cite{I-K}): it is a one-parameter family of maps $\Lambda_t : \mathcal{T}(\mathcal{H}) \rightarrow \mathcal{T}(\mathcal{H})$ satisfying
\begin{enumerate}
  \item $\Lambda_t$ is a positive map, i.e.\ $\Lambda_t : \mathcal{T}_+(\mathcal{H}) \rightarrow \mathcal{T}_+(\mathcal{H})$,
  \item $\Lambda_t $ is strongly continuous,
  \item $\Lambda_t \Lambda_s = \Lambda_{t+s}$, for all $t,s \geq 0$ (Markov property).
\end{enumerate}
Now, the celebrated Hille-Yosida theorem \cite{H-J} states that there exists a densely defined generator
\begin{equation}\label{L}
  \mathcal{L}\rho \;=\; \lim_{t \rightarrow 0} \frac 1 t (\Lambda_t \rho - \rho) \,,
\end{equation}
such that
\begin{equation}\label{ME}
  \dot{\Lambda}_t \;=\; \mathcal{L} \Lambda_t \,,
\end{equation}
together with the initial condition $\lim_{t \rightarrow 0} \Lambda_t\rho = \rho$. The problem is to characterize the properties of $\mathcal{L}$ such that the solution of (\ref{ME}) defines a dynamical semigroup $\Lambda_t$. This problem is still open nowadays.

In their two seminal papers \cite{GKS,Lindblad}, GKLS considered a restricted problem: first of all they focused on a \emph{subclass\/} of positive maps --- completely positive maps (see Sect.\ \ref{SECTION-CP}). Moreover,
\begin{enumerate}
  \item GKS considered only finite-dimensional Hilbert spaces, i.e.\ ${\rm dim}\, \mathcal{H} =N < \infty$.
  \item Lindblad considered the infinite-dimensional case, but replaced strong continuity by uniform continuity. This considerably simplifies the analysis, since uniform continuity implies that the corresponding generator (\ref{L}) defines a bounded operator (see \cite{Werner} for the analysis of unbounded generators).
\end{enumerate}
A few technical details will help us clarify. Let $\mathbb{M}_N$ denote the algebra of $N \times N$ complex matrices. GKS worked in the Schr\"odinger picture and arrived at the following result:

\begin{Theorem}[Theorem 2.2 in \cite{GKS}]
A linear operator $\mathcal{L} : \mathbb{M}_N \rightarrow \mathbb{M}_N$ is the generator of a completely positive semigroup of $\mathbb{M}_N$ if it can be expressed in the following form
\begin{equation}
\label{GKS}
  \mathcal{L} \rho \;=\; -i [H,\rho] + \frac 12 \sum_{k,l=1}^{N^2-1} c_{kl} \Big( [F_k,\rho F_l^*] +  [F_k\rho, F_l^*]  \Big) \,,
\end{equation}
where $H=H^*$, ${\rm tr}H=0$, ${\rm tr}F_k=0$, ${\rm tr}(F_k F^*_l)=\delta_{kl}$, $k,l= 1,2,\ldots,N^2-1$,
and $[c_{kl}]$ is a complex positive
matrix.
\end{Theorem}
We shall call (\ref{GKS}) the GKS form of the GKLS generator and $[c_{kl}]$ the Kossakowski matrix.
Given a basis $F_k$, the generator is fully determined by the Hamiltonian $H$ and the Kossakowski matrix.

On the other hand, Lindblad worked in the Heisenberg picture at the level of $\mathfrak{B}(\mathcal{H})$ ---
the algebra of bounded operators on $\mathcal{H}$. Using completely different techniques he arrived at
\begin{Theorem}[Theorem 2 in \cite{Lindblad}]
A linear operator $\mathcal{L}^* : \mathfrak{B}(\mathcal{H}) \rightarrow \mathfrak{B}(\mathcal{H})$ is the generator of a completely positive semigroup if and only if
\begin{equation}
\label{Lind1}
  \mathcal{L}^* X \;=\; i [H,X] +  \sum_j \Big( V^*_j X V_j - \frac 12\{ V^*_j V_j, X\} \Big) \,,
\end{equation}
where $V_j$, $\sum_j V_j^* V_j \in \mathfrak{B}(\mathcal{H})$, $\{A,B\}= AB + BA$, and $H$ is a self-adjoint element in $\mathfrak{B}(\mathcal{H})$. The corresponding generator in the Schr\"odinger picture is of the form
\begin{equation}
\label{Lind}
  \mathcal{L}\rho \;=\; -i [H,\rho] +  \frac 12 \sum_j \Big(  [V_j,\rho V^*_j] +  [V_j\rho, V^*_j]\Big) \,.
\end{equation}
\end{Theorem}
We shall call (\ref{Lind}) the Lindblad form of the GKLS generator. Observe that in (\ref{Lind}) the choice of $H$ and $V_j$ is not unique,  whereas in (\ref{GKS}), after fixing the basis $F_l$, the traceless Hamiltonian is uniquely defined.
\begin{Remark}\rm
By GKLS (or standard) form of the generator one usually understands
\begin{equation}
\label{GKSL}
  \mathcal{L}\rho \;=\;  -i [H,\rho] + \Phi \rho - \frac 12 \{ \Phi^* \mathbb{I},\rho\} \,,
\end{equation}
where $H$ is a self-adjoint operator in the Hilbert space of the system $\mathcal{H}$, $\Phi$ is a completely positive map, and $\Phi^*$ stands for its dual (Heisenberg picture). Equivalently, representing
\begin{equation}
  \Phi \rho \;=\; \sum_j V_j \rho V_j^* ,
\end{equation}
one gets
\begin{equation}
\label{GKSL-2}
  \mathcal{L}\rho \;=\;  -i [H,\rho] + \frac 12 \sum_j (2V_j  \rho V_j^* - V_j^* V_j \rho - \rho V_j^* V_j)\, .
\end{equation}
In the unbounded case there are examples of generators which are not of the standard form (cf.\ \cite{Werner}).
\end{Remark}
This is what GKLS obtained in 1975--1976. A few years later, Alicki and Lendi \cite{Alicki} published a monograph, presenting both the theory and physical applications of Markovian semigroups. Modern monographs include Weiss~\cite{Weiss}, Breuer and Petruccione \cite{Breuer}, and Rivas and Huelga \cite{Rivas}.\footnote{Interestingly, R. Alicki, H.-P. Breuer, F. Petruccione and S. Huelga were among the participants of the 48th SMP in Toru\'n.}

\section{Master (Kinetic) Equations before GKLS}

Master equations --- also known as kinetic equations --- were used to describe dissipative phenomena long before the GKLS articles. Let us briefly review their interesting history.
It is of great interest to analyze the structure of these equations, that in some cases is very similar to GKLS. In general (but not always), physicists cared about positivity and trace-preservation.
Nobody knew (and bothered) about complete positivity before GKLS.

\subsection{Landau's approach to the damping problem}

Already in 1927 Landau \cite{Landau} derived a kinetic equation for the elements of the density matrix of the radiation field interacting with charged matter (in the dipole approximation). Remarkably, in the same article Landau introduced for the first time the concept of density matrix (that would be defined in a more formal way the same year by von Neumann \cite{Neumann}).  In modern language the Landau equation may be (re)written as follows
\begin{equation}
\label{Landaueq}
  \dot{\rho} \;=\; \frac 12 \gamma ( 2a\rho a^* -  \rho a^*a -  a^*a\rho) \,,
\end{equation}
where $a$ and $a^*$ are the annihilation and creation operators for the radiation field, and $\gamma > 0$ is a damping constant. This is the correct form of the master equation. Notice that  $a$ and $a^*$ are unbounded.

\subsection{Optical potential}
\label{sec-nHH}

In optics, the interaction between light and a refractive and absorptive medi\-um is described by introducing a complex refractive index, whose imaginary component absorbs light. In close analogy,
during the golden years of nuclear physics, the so-called ``optical potentials" were often used to describe the coherent scattering of slow neutrons travelling through matter 
[14--17].
The scattering and absorption of nucleons (such as neutrons) by nuclei was treated by averaging effective neutron-nucleus interaction potentials over many nuclei, to yield a neutron-matter (complex) optical potential.

We use in the following modern language. Let $V \geq 0$ be a non-negative operator (``optical" potential). The non-Hermitian ``Hamiltonian"
\begin{equation}
H' \;=\; H - i V\,,
\label{eq:HiV}
\end{equation}
yields the evolution
\begin{equation}
\psi \;=\; \e^{(-i H - V)t} \psi_0\,,
\label{eq:HiV-1}
\end{equation}
where $\psi_0$ is the initial wave function and $V$ is switched on at some instant of time or some position in space (when the particle impinges on the nucleus). The density matrix evolves according to
\begin{equation}
\label{optical}
  \dot{\rho} \;=\; -i [H,\rho] - (V \rho + \rho V) \,.
\end{equation}
The trace is not preserved and probabilities are not conserved (particle absorption). Notice that if $V$ were not positive-defined, probabilities would be allowed to become larger than 1 (particle creation).
Optical potentials have played an important role in nuclear physics.

\subsection{Lamb equation}

Lamb, analyzing the theory of the optical maser \cite{Lamb}, considered the following equation for the (unnormalized) density matrix $\rho$ of a 2-level atom (equation (18) in \cite{Lamb}):
\begin{equation}
\label{lamb}
  \dot{\rho} \;=\; -i[H,\rho] - \frac 12 (\Gamma \rho+ \rho \Gamma) \,,
\end{equation}
where $\Gamma$ is a diagonal matrix with positive diagonal elements $\gamma_1$ and $\gamma_2$, the decay constants of the two states of the atom. This equation generates the legitimate evolution
\begin{equation}
  \rho_0 \longrightarrow \rho_t \;=\; e^{(-iH - \frac 12 \Gamma)t} \rho_0 e^{(iH - \frac 12 \Gamma)t} \,,
\end{equation}
but it is not trace-preserving. Like with an optical potential (preceding subsection) positivity of $\Gamma$ implies ${\rm tr} \dot{\rho} < 0$.

Compare (\ref{optical}) and (\ref{lamb}) with (\ref{GKSL-2}): the (jump) terms $2V \rho V$ and $\Gamma \rho \Gamma$ are missing. It is clear that the Lamb equation (\ref{lamb}) may be ``cured" by adding to the r.h.s.\ an additional term ``$\Phi \rho$", where $\Phi$ is an arbitrary completely positive map such that $\Phi^* \mathbb{I} = \Gamma$. In this way, Lamb's becomes a legitimate GKLS equation.
Similarly for (\ref{optical}).

\subsection{Redfield equation}

Redfield, using the Born-Markov approximation, derived the following Markovian master equation \cite{Redfield}
\begin{equation}
\label{redf}
  \dot{\rho} \;=\; -i[H,\rho] - \sum_\alpha [V_\alpha, X_\alpha \rho - \rho X_\alpha^\dagger] \,,
\end{equation}
where the operators $V_k$ are defined via the system--bath interaction Hamiltonian $H_I = \sum_\alpha V_\alpha \ot B_\alpha$, and
\begin{equation}
\label{redf2}
 X_\alpha \;=\; \sum_\beta \int\limits_0^\infty h_{\alpha\beta}(\tau) \widetilde{V}_\beta(-\tau) d\tau \,,
\end{equation}
$h_{\alpha\beta}(t) = {\rm tr}( \widetilde{B}_\alpha(t) B_\beta \rho_B)$ being the two-point bath correlation function,
$\rho_B$ the initial state of the bath, and tilde stands for the interaction picture. It is obvious that ${\rm tr}\rho_t = {\rm tr}\rho_0$. However, the positivity of $\rho_t$ is  not guaranteed (cf.\ the detailed discussion in \cite{Fabio}).

\subsection{Quantum optics}

Master equations of the form (\ref{master}) for the density operator $\rho$, with $\mathcal{L}$ having the form of GKLS generator (\ref{L})--(\ref{GKS}), were used in quantum optics long before the GKLS articles \cite{GKS,Lindblad}. For example, the Stuttgart group (Weidlich, Haake, Risken and Haken)
[21--23],
working on the theory of the laser, derived the following equation for the statistical operator $R$ of the laser field
\begin{equation}\label{R}
  \dot{R} \;=\; - i[H,R] + \Big( \frac{\partial R}{\partial t} \Big)_{\rm incoh}\, ,
\end{equation}
where the {\em incoherent} term is given by
\begin{equation}
  \Big( \frac{\partial R}{\partial t} \Big)_{\rm incoh} \;=\;
  \nu ([b,R b^*] +  [bR, b^*]) + \delta ([b^*,R b] +  [b^*R, b]) \,,
\end{equation}
$b$ and $b^*$ are the annihilation and creation operators of the field mode,  $\nu$ is determined by the coupling between the laser field and the atoms, and
\begin{equation}
  \delta \;=\; e^{-\Omega \hbar/kT} \nu \,,
\end{equation}
with $T$ being the temperature of the atoms.  Moreover, the same authors recovered the well-known formula for the mean number of thermal photons
\begin{equation}
  n_{\rm th}(T)\;=\; \frac{\delta}{\nu - \delta} \;=\; \frac{1}{e^{\Omega \hbar/kT}-1} \,.
\end{equation}
This proves that the master equation (\ref{R}) properly describes the interaction of the laser field with atomic matter. Clearly, (\ref{R}) has exactly the form of  (\ref{GKSL-2}), with $V_1 = \sqrt{2 \nu}\,b$ and $V_2 = \sqrt{2 \delta}\,b^*$, showing that the GKLS master equation worked perfectly well ten years before GKLS's papers \cite{GKS,Lindblad}. This research is summarized in Haken's book \cite{Haken-book} and the review papers by Haake \cite{Haake} and Agarwal \cite{Agarwal}. The same equation was independently used at the same time by Shen \cite{Shen} and the slightly isolated Russian school of Belavin \emph{et al.} \cite{Belavin} and Zel'dovich \emph{et al.} \cite{Popov} (this list is by no means exhaustive).

\subsection{Davies approach to Markovian Master Equations}

In a series of papers  [30\,--\,33]
(see also the book \cite{DAVIES}) Davies presented a general approach to Markovian master equations (\ref{ME}) using the so-called weak-coupling limit.
Based on the Nakajima-Zwanzig projection techniques \cite{Nakajima,Zwanzig} (see also \cite{KTH,GZ}) and van Hove's idea of time rescaling \cite{van-Hove}, Davies derived the following form of the generator (in the interaction picture)
\begin{eqnarray}\label{L-D}
  \mathcal{L}\rho &=& \sum_\omega \sum_{\alpha,\beta} \Big\{ -is_{\alpha\beta}(\omega)[V_\beta^\dagger(\omega) V_\alpha(\omega),\rho]\nonumber  \\
  &&+\; \frac 12 \widehat{h}_{\alpha\beta}(\omega)( [V_\alpha(\omega)\rho,V_\beta^\dagger(\omega)] + [V_\alpha(\omega),\rho V_\beta^\dagger(\omega)] )  \Big\} \,,
  \end{eqnarray}
where $\omega$ are the Bohr frequencies of the system Hamiltonian, $\widehat{h}_{\alpha\beta}(\omega)$ denotes the Fourier transform of the two-point correlation function, and
\begin{equation}
  s_{\alpha\beta}(\omega) \;=\; \frac{1}{2\pi} \mathcal{P}\!\! \int\limits_{-\infty}^{\infty} \frac{\widehat{h}_{\alpha\beta}(\xi)}{\xi - \omega}d\xi \,,
\end{equation}
with $\mathcal{P}$ denoting the principal part. Bochner's theorem \cite{RS2} implies that the
Kossakowski matrix $c_{\alpha\beta} = \frac 12 \sum_\omega \widehat{h}_{\alpha\beta}(\omega)$ is positive definite and hence the generator defined by (\ref{L-D}) is a legitimate GKLS generator.
It  is usually called a Davies generator and many physical systems are properly described by (\ref{L-D}).

\subsection{Quantum stochastic processes}

In the '70s there was a parallel activity on quantum stochastic processes. This topic is closely related to the notion of complete positivity and has a nontrivial overlap with the theory of master equations developed by GKLS. For a historical review of this interesting activity, see \cite{Accardi}.

\section{Complete Positivity and its Appearance in Phy\-sics}
\label{SECTION-CP}

Let us now look at a central hypothesis in the derivation of a quantum master equation. The very concept of complete positivity was introduced by Stinespring in a seminal paper \cite{CP}. Let us recall that a linear map $\Phi$ between two $C^*$-algebras $\mathfrak{A}$ and $\mathfrak{B}$ is positive if $\Phi a \geq 0$ for each $a \geq 0$ ($a \geq 0$ if $a = xx^*$ for some $x \in \mathfrak{A}$).
Now, one calls a linear map $\Phi$ $k$-positive if the $k$-amplification
\begin{equation}
  \oper_k \ot \Phi : \mathbb{M}_k \ot \mathfrak{A} \longrightarrow \mathbb{M}_k \ot \mathfrak{B} \,,
\end{equation}
is positive, $\mathbb{M}_k$ being the algebra of $k \times k$ complex matrices. Finally, $\Phi$ is completely positive (CP) if it is $k$-positive for all $k \in \mathbb{N}$. A remarkable result due to Stinespring consists in the following
\begin{Theorem}
Let $\mathfrak{A}$ be a unital $C^*$-algebra and let $\Phi :  \mathfrak{A} \rightarrow  \mathfrak{B}(\mathcal{H})$ be a CP map. Then there exists a Hilbert space $\mathcal{K}$, a unital $*$-homomorphism $\pi : \mathfrak{A} \rightarrow  \mathfrak{B}(\mathcal{K})$, and  a bounded operator
$V : \mathcal{H} \rightarrow \mathcal{K}$, with $||\Phi I|| = ||V||^2$, such that
\begin{equation}
  \Phi a \;=\; V \pi(a) V^* ,
\end{equation}
for all $a \in \mathfrak{A}$.
\end{Theorem}
The notion of CP maps slowly appeared in the mathematical literature in the '60s. Nakamura, Takesaki and Umegaki \cite{Nakamura} proved that ``\textit{the conditional expectation in a not necessarily commutative probability is completely positive in the sense of Stinespring}''. In his review on positive maps \cite{Stormer} St{\o}rmer included completely positive maps and introduced the notion of positive decomposable map (see also \cite{Stormer-74} and the recent monograph \cite{STORMER}). In a seminal paper \cite{Woronowicz},
Woronowicz showed that all maps $\Phi : M_n \to M_m$ with $mn \leq 6$, if positive, are automatically decomposable.
Another seminal paper developing the concept of completely positive  maps is due to Arveson \cite{Arveson}, who introduced the notions of completely bounded, completely contractive and completely isometric maps (see the recent monograph \cite{Paulsen}).

In physics the importance of completely positive maps was first recognized by the German school of mathematical physics (Ludwig, Haag, Kraus, Hellwig and others) [50\,--\,53]. 
In his seminal paper \cite{G5}, Kraus introduced the notion of quantum operation $\Phi$. Let us quote from \cite{KRAUS}:
\begin{quotation}\noindent
$\ldots$ {\em for physical reason the mapping $\Phi$  must have still another property, called {\rm complete positivity}  and being somehow stronger than ``ordinary'' positivity, such that the mapping $\Phi$ is given by}
\end{quotation}
\begin{equation}
\label{KAK}
  \Phi A \;=\; \sum_{\alpha}^r K_\alpha A K_\alpha^* .
\end{equation}

This is the celebrated Kraus representation. Interestingly, a few years later Choi \cite{Choi1,Choi2} derived (\ref{KAK}) without the  knowledge of Kraus' work (GKLS cite both Kraus and Choi). Choi observed that in a finite-dimensional setting the map $\Phi$ is completely positive if and only if the following (so-called Choi) matrix
\begin{equation}
  \mathcal{C} \;=\; (\oper_n \ot \Phi) P^+_n \,,
\end{equation}
is positive definite ($P^+_n$ is a projector onto the maximally entangled state in $\mathbb{C}^n \ot \mathbb{C}^n$, and $\oper_n$ is the  identity map in $\mathbb{M}_n$).
Remarkably, complete positivity reduces to the positivity of $\oper_n \ot \Phi$ on a single projector $P^+_n$. It is therefore clear that complete positivity introduces a great simplification for the analysis of dynamical maps and the corresponding master equations.

How did GKLS learn about complete positivity? The Advanced Study Institute on ``Foundations of Quantum Mechanics and Ordered Linear Spa\-ces" was held in Marburg between March 26th and April 6th, 1973.
In the introduction to the Proceedings of this meeting one reads:
\begin{quotation} \noindent
\textit{Mathematicians and physicists participated in the meeting. The lectures of the Institute were intended to prepare a common basis for discussion between mathematicians and physicists for the future research and on foundations of quantum mechanics by ordered linear spaces. A series of lectures (``Course") provided a coherent introduction into the field of ordered normed vector spaces and their application to the foundation of quantum
mechanics. Additional lectures treated special mathematical and physical topics, which were in more or less connection to the lectures of the course.}
\end{quotation}
In the list of mathematical lectures one finds a lecture by E.\ St{\o}rmer on ``\textit{Positive linear maps of $C^*$-algebras}". The list of physics lectures contains those of G.\ Ludwig on ``\textit{Measuring and preparing processes}" and K.\ Kraus on  ``\textit{Operations and effects in the Hilbert space formulation of quantum theory}".
V.\ Gorini contributed with the talk ``\textit{Irreversibility and dynamical maps of statistical operators}". After the Marburg meeting Gorini went to Texas and spread the good news of complete positivity to Kossakowski and Sudarshan. He then met Lindblad in Stockholm in January 1975 (see Lindblad's recollection in Sect.\ \ref{Lrec}). Lindblad was already aware of this concept, having used it in his 1974 proof of a fundamental property of quantum relative entropy \cite{Lindbald-75} (quantum relative entropy is monotonic with respect to completely positive trace preserving maps, called nowadays quantum channels).

\section{Positivity vs.\ Complete Positivity --- the Great Simplification}

Due to Stinespring's theorem, completely positive maps are fully characterized.
The problem of the characterization of positive maps is still open and is equivalent to the characterization of the entangled states of bipartite quantum systems \cite{QIT,HHHH}. In the beginning of the '70s Kossakowski (still without the knowledge of complete positivity) addressed the problem of finding the necessary and sufficient conditions for a
bounded generator $\mathcal{L}$ of the semigroup of positive maps $\Lambda_t = e^{t \mathcal{L}}$ for $t \geq 0$. Note that if $\Lambda_t$ is trace-preserving and hermiticity-preserving (i.e.\ $\Lambda_t X^* = (\Lambda_t X)^*)$, then any (bounded) generator must have the standard form (\ref{GKSL})
\begin{equation}
  \mathcal{L}\rho \;=\;  -i[H,\rho] + \Phi \rho - \frac 12 \{ \Phi^* \mathbb{I},\rho\} \,,
\end{equation}
with $\Phi$ a hermiticity-preserving map. GKLS proved that the corresponding dynamical map $e^{t \mathcal{L}}$ is completely positive iff the map $\Phi$  is completely positive as well.
What happens if $e^{t \mathcal{L}}$ is \textit{only positive}? In such a case, the positivity of $\Phi$ is only sufficient, but not necessary!
Kossakowski proved the following \cite{AK-1972}
\begin{Theorem}\label{TH-72}
A bounded linear operator $\mathcal{L}$ on  $\mathcal{T}(\mathcal{H})$ generates a dynamical semigroup $\Lambda_t$ if and only if for every resolution of identity $P_i P_j = \delta_{ij} P_i$, ${\rm tr} P_i < \infty$, and  $\sum_i P_i = \mathbb{I}_\mathcal{H}$ one has
\begin{equation}\label{PLP}
  {\rm tr}( P_i \mathcal{L} P_j ) \geq 0  \,, \quad i \neq j \,,
\end{equation}
and $\sum_i  {\rm tr}( P_i \mathcal{L} P_j ) = 0$.
\end{Theorem}
Unfortunately, the general structure of $\mathcal{L}$ satisfying (\ref{PLP}) is not known. It is clear that if ${\rm dim}\, \mathcal{H} = N < \infty$, then $\mathcal{L}$ has the structure defined by (\ref{GKS}), but there is no simple condition for the matrix $[c_{kl}]$.  As an example \cite{AK-1}
consider $N=2$ and
\begin{equation}
  \mathcal{L}\rho \;=\; \sum_{k=1}^3 \gamma_k (\sigma_k \rho \sigma_k - \rho) \,,
\end{equation}
that is, the Kossakowski matrix $c_{kl} = \gamma_k \delta_{kl}$ is diagonal. $\mathcal{L}$ is a GKLS generator iff $\gamma_k \geq 0$, that is, the map
$  \Phi\rho = \sum_{k=1}^3 \gamma_k \sigma_k \rho \sigma_k\,$ is completely positive.  However, $\mathcal{L}$ satisfies (\ref{PLP}) iff
\begin{equation}
\label{exex}
  \gamma_1 + \gamma_2 \geq 0 \, , \quad  \gamma_2 + \gamma_3 \geq 0 \, , \quad  \gamma_3 + \gamma_1 \geq 0 \, .
\end{equation}
These conditions, which are much weaker than $\gamma_k \geq 0$, do not guarantee the positivity of the map $\Phi$.
For example, by taking $\gamma_1=\gamma_2 = -\gamma_3=1$ one finds $\Phi \rho = \sigma_1 \rho \sigma_1 + \sigma_2 \rho \sigma_2 - \sigma_3 \rho \sigma_3$
which is not positive, although the $\gamma$'s satisfy the conditions (\ref{exex}). It is therefore clear that the positivity of $\Phi$ is not necessary for the positivity of the dynamical map $e^{t \mathcal{L}}$.

A similar problem was addressed in 1969 by Belavin \emph{et al.}~\cite{Belavin}. They analyzed the following kinetic equation (equation (14) in \cite{Belavin})
\begin{equation}
  \dot{\rho} \;=\; \frac 12 \sum_{i,j=1}^N \gamma_{ij} ( 2 A_i \rho A_j^\dagger - A_j^\dagger A_i \rho - \rho  A_j^\dagger A_i  ) \,.
\end{equation}
Clearly, being not aware of complete positivity, they simply looked for the evolution
$\rho_0 \rightarrow \rho_t = \Lambda_t \rho_0$ realized by a family of positive maps $\Lambda_t$.
If $N=1$ they showed that $\rho_t$ is a legitimate density matrix for all $t  \geq 0$ iff $\gamma > 0$ (in this case there is only a single $\gamma$). Hence for a single dissipation/decoherence channel the correct form of the generator was already known in 1969!
The authors claimed: {\em it is not difficult to generalize the proof given here to the case  $N>1$}.
The answer they provided (without proof) reads: the matrix $[\gamma_{ij}]$ has to be positive definite.
Clearly, this is very intuitive, but  as we know this condition is both necessary and sufficient for complete positivity, but it is not necessary for the simple positivity required in \cite{Belavin}.
This example shows that the condition for the positivity of the Kossakowski matrix $[\gamma_{ij}]$ was anticipated as a natural generalization of positivity for a single dissipative channel.
However, only GKLS proved the consistency of this condition with the \emph{complete\/} positivity of the dynamical map. The spirit of the article by Belavin {\it et al.}~\cite{Belavin} is nonetheless very close to that of the GKLS papers. It is unfortunate that their work  is not well known.\footnote{The article by Belavin {\it et al.}~\cite{Belavin} has been recently cited in \cite{PRX}. We wonder if this is because one of the authors -- Victor Albert -- participated in the 48th SMP. If our guess is correct, this proves that the SMP meetings can be particularly useful for young and promising physicists!}

In his recent book, Weinberg \cite{Weinberg} wrote: {\it The Lindblad equation can be derived as a straightforward application of an earlier result by A.\ Kossakowski \cite{Kossa1972}, eq.~(77)}.
This interesting statement can be considered correct or incorrect (depending on the viewpoint and level of rigor one adopts) and shows how delicate questions of priority can be.
In his 1972 paper Kossakowski wrote the right equation using only positivity, and from those premises one cannot obtain the GKLS equation, whose proof requires complete positivity.

\section{The Mystery of Sudarshan-Mathews-Rau Paper}

In 1961 Sudarshan, Mathews, and Rau  \cite{SMR} published a remarkable paper entitled {\em Stochastic Dynamics of Quantum Mechanical Systems}.
In the abstract the authors state:
\begin{quotation} \noindent
{\em The most general dynamical law for a quantum mechanical system with a finite number of levels is
formulated. A fundamental role is played by the so-called ``dynamical matrix" whose properties are stated
in a sequence of theorems. A necessary and sufficient criterion for distinguishing dynamical matrices corresponding
to a Hamiltonian time-dependence is formulated}.
\end{quotation}
Sudarshan, Mathews, and Rau analyzed the evolution of the density matrix represented by the following linear relation
\begin{equation}
  \rho_{rs}(t) \;=\; \sum_{r',s'} A_{rs,r's'}(t,t_0) \rho_{r's'}(t_0)
\end{equation}
and found that $\rho_{rs}(t)$ defines a density matrix for $t > t_0$ if and only if the $A$ matrix satisfies
the following properties:
\begin{eqnarray}\label{A}
  A_{sr,s'r'} = \overline{A_{rs,r's'}} \, , &&\quad \mbox{(Hermiticity)} \nonumber \\
 \sum_{r,s,r',s'}\overline{x}_r x_s A_{rs,r's'} y_{r'} \overline{y_{s'}} \;\geq\; 0 \, ,&&\quad
 \mbox{(positivity)} \\
 \sum_r A_{rr,r's'} = \delta_{r's'} \, . && \quad \mbox{(trace-preservation)}\,. \nonumber
\end{eqnarray}
If one knows how to represent a matrix satisfying the above properties, the problem is solved. However, as the authors remarked, these conditions are {\em fairly complicated}. In order to solve the problem they proposed to analyze the matrix $B$ defined by
\begin{equation}
  B_{rr',ss'}  := A_{rs,r's'}  .
\end{equation}
In modern language, $B$, which was named {\em dynamical matrix\/} in \cite{SMR}, is nothing but the realignment of $A$ (see e.g.~\cite{Karol}). Now, the authors claimed (without proof) that properties (\ref{A}) lead to the following properties of $B$
\begin{eqnarray}\label{B}
 B_{rr',ss'} = \overline{B_{ss',rr'}} \, , &&\quad \mbox{(Hermiticity)}  \nonumber \\
 \sum_{r,s,r',s'} \overline{z}_{rr'} B_{rr',ss'} z_{ss'} \;\geq\; 0\,, &&\quad \mbox{(positivity)} \label{B-1} \\
 \sum_r B_{rr',rs'} = \delta_{r's'} \, , && \quad \mbox{(trace-preservation)}\, .  \nonumber
\end{eqnarray}
Finally, they derived the remarkable result that $B$ satisfies (\ref{B}) if and only if there exists a set of parameters $\mu_\alpha \geq 0$, and a collection of $n \times n$ complex matrices $W_\alpha$, such that
\begin{equation}
  B_{rr',ss'} \;=\; \sum_{\alpha=1}^{n^2} \mu_\alpha (W_\alpha)_{rr'}   \overline{(W_\alpha)_{ss'}} \,.
\end{equation}
In such a case, the evolution of $\rho$ is realized via
\begin{equation}\label{KRAUS}
  \rho \;\longmapsto\; \sum_{\alpha=1}^{n^2} \mu_\alpha W_\alpha  \rho W_\alpha^\dagger \,,
\end{equation}
where $\mu_\alpha \geq 0$, and $\sum_{\alpha=1}^{n^2} \mu_\alpha W_\alpha^\dagger  W_\alpha = \mathbb{I}$. This is nothing but the celebrated Kraus representation
(\ref{KAK}) of a quantum channel, derived almost \emph{ten years\/} before Kraus \cite{G5}!
How was it possible to derive the Kraus representation without making any use of completely positive maps? Note that in passing from (\ref{A}) to (\ref{B}) there is a gap.
The second condition for $B$ should read
\begin{equation}
\label{B-2}
\sum_{r,s,r',s'} \overline{x}_r x_s B_{rr',ss'} y_{r'} \overline{y_{s'}} \;\geq\; 0\,.
\end{equation}
This condition states that the dynamical matrix $B$ is not necessarily positive but only block-positive (cf.~\cite{Karol}). This was a prophetic error, anticipating the Kraus representation of a quantum channel.
It is clear that this remarkable paper was written too early and the community in 1961 was not prepared to grasp its elegance and predictive power!

In modern language, $B$ is nothing but the Choi matrix corresponding to the dynamical map, and condition
(\ref{B-1})
is, as Choi proved in 1975 \cite{Choi2}, equivalent to complete positivity. The weaker condition (\ref{B-2})
was proved by Jamio{\l}kowski \cite{JAM} to be equivalent to the positivity of the map.
The correspondence ``map  $\longleftrightarrow$  $B$-matrix'' is usually called Choi-Jamio{\l}kowski isomorphism. Actually, such correspondence and its properties were analyzed by de Pillis \cite{Pillis}
in 1967 (Kossakowski cited \cite{Pillis} in his 1972 paper \cite{Kossa1972}, in connection with semi-groups of positive maps). See also the recent review \cite{Topical}.
Clearly, this correspondence was already noted in \cite{SMR}.

\section{Recollection from G\"oran Lindblad}
\label{Lrec}

Unfortunately,  G\"oran Lindblad was unable to participate in the 48th SMP meeting in Toru\'n.
However, he sent the following ``recollection slide" to the organizers:
\begin{quotation} \noindent
{\em In the Spring of 1974 I received a letter from Professor Ingarden (dated June 22) inviting
me to take part and give a seminar on entropy and related problems (the subject of my
recent thesis). I answered that I would be glad to come, but that I would prefer to talk
about some even more recent results on quantum dynamical semigroups. I had only just
realized that the CP condition makes the structure so much simpler.
In fact, after passing my thesis in May 1974, I went back to my desk to consider the
problem, and after a week or so I realized that I knew how to make an `if and only if
proof'. I was quite euphoric for a week or two, and then considered how and where to
publish.

When I got Prof.\ Ingarden's letter, I decided to use the conference for a first
announcement.

After my talk (December 5) Prof.\ Ingarden told me that he had received news from
Texas of similar results. Later, in January 1975, Vittorio Gorini passed through Stockholm
on his way back home, and we could compare things and agree that there was a good
deal of overlap, in fact the results were essentially the same, save for the mathematical
machinery.}
\end{quotation}
\def\arraystretch{1.4}
\begin{table}[t]
\centering{}%
\begin{tabular}{|c|l|}
\hline
Date & \hspace*{25mm} Event \\
\hline\hline
1955 & Stinespring publishes \cite{CP} \tabularnewline
\hline
1961 & Sudarshan, Mathews, and Rau publish \cite{SMR}  \tabularnewline
\hline
1971 & Kraus publishes \cite{G5} \tabularnewline
\hline
1972 & Kossakowski publishes \cite{Kossa1972}  \tabularnewline
\hline
26 March\,--\,6 April 1973 & Gorini attends Marburg conference, where \\ & St{\o}rmer and Kraus mention complete \\ & positivity \tabularnewline
\hline
May 1974 & Lindblad submits \cite{Lindbald-75} \tabularnewline
\hline
September\,--\,December 1974 & Gorini and Kossakowski visit Sudarshan in \\ & Texas \tabularnewline
\hline
December 1974 & Lindblad visits Ingarden in Toru\'n \tabularnewline
\hline
January 1975 & Gorini visits Lindblad in Stockholm \tabularnewline
\hline
1975 & Choi publishes \cite{Choi2} \tabularnewline
\hline
March and April 1975 & GKLS articles \cite{GKS,Lindblad} are submitted \tabularnewline
\hline
1980 & Kraus spends sabbatical year at University \\ & of Texas at Austin; Sudarshan, Wheeler,  \\ &   A.\ B\"ohm and Wootters are there
\tabularnewline
\hline
1983 & Kraus publishes \cite{KRAUS}
\tabularnewline
\hline
\end{tabular}\caption{Chronology}
\label{table:chrono}
\end{table}
\def\arraystretch{1}

\section{Conclusions and Outlook}

In retrospect, there were a number of concomitant factors that led to the precise mathematical formulation of the evolution equations for open quantum dynamical systems. The chronology of some of these episodes is summarized in Table \ref{table:chrono}.
We leave to the wisdom of Noam Chomsky \cite{Chomsky} the final judgement on the ``level of significance" at which there are differences and similarities.

In our humble opinion, there are many similarities, and also a few differences, among the seminal articles discussed in this note.
We venture to suggest that (\ref{master}), with (\ref{GKS}), (\ref{Lind}) and (\ref{GKSL}), be named ``Gorini-Kossakowski-Lindblad-Sudarshan (GKLS) equation"
and (\ref{superoperator}) the ``Kraus-Stinespring-Sud\-ar\-shan (KSS) representation".

\section*{Acknowledgments}
D.\,C. was partially supported by the National Science Centre project number 2015/17/B/ST2/02026.
S.\,P. was partially supported by INFN through the project ``QUANTUM".
S.\,P. would like to thank the organizers of the
48th Symposium on Mathematical Physics (Toru\'n, June 10--12, 2016) for the invitation and kind hospitality.

\end{document}